\documentclass{elsart}
\usepackage{graphicx}
\usepackage{color}
\usepackage{lineno}

\usepackage{amssymb}

\newcommand \beq{\begin{eqnarray}}
\newcommand \eeq{\end{eqnarray}}
\newcommand{\bcen}{\begin{center}}
\newcommand{\ecen}{\end{center}}
\newcommand{\bfig}{\begin{figure}}
\newcommand{\efig}{\end{figure}}
\newcommand{\degr}{$\,^{\rm o}$}
\def\agev{~A$\cdot$GeV}

\begin{document}

\begin{frontmatter}


\title{Laser calibration system for the CERES Time Projection Chamber}


\author[1]{Dariusz Mi\'{s}kowiec\corauthref{cor1}}
\ead{D.Miskowiec@gsi.de}
\ead[url]{http://www.gsi.de/{\~{}}misko}
\corauth[cor1]{corresponding author}
\author[1,2]{Peter Braun-Munzinger}

\address[1]{Gesellschaft f{\"u}r Schwerionenforschung mbH, Darmstadt, Germany}
\address[2]{Institut f\"ur Kernphysik, TU Darmstadt, Germany}

\title{}


\begin{abstract}
A Nd:YAG laser was used to simulate charged particle tracks at known 
positions in the CERES Time Projection Chamber at the CERN SPS. 
The system was primarily developed to study the response of the readout 
electronics and to calibrate the electron drift velocity. 
Further applications were the determination of the gating grid 
transparency, the chamber position calibration, and long-term monitoring 
of drift properties of the gas in the detector. 
\end{abstract}

\end{frontmatter}

\newpage

\section{Introduction}
\label{sec_intro}
UV lasers have been widely used to simulate charged particle 
tracks in gaseous detectors~\cite{guo82,kon83,ame85,hil86}. 
Via the two-photon absorption mechanism the laser beam ionizes impurities 
in the gas and produces a signal which, with proper adjustment of the 
intensity, imitates that of a straight particle track. 
Large drift chambers at LEP~\cite{delphi,aleph,opal} 
and at the SPS~\cite{na35} made extensive use of this method. 
Similarly, the large cylindrical Time Projection Chamber (TPC)~\cite{tpc-nim}, 
added to the setup of the CERES experiment at the CERN SPS in 1998, 
was equipped with a laser system for calibration and monitoring purposes. 
In this paper we present the layout and the components of this system, 
describe the procedures employed to determine the absolute position of 
the laser beam, and show results of the calibration. 

\section{Layout of the laser system}
\label{sec_layout}
\begin{figure}[t]
\bcen
\includegraphics[width=14cm]{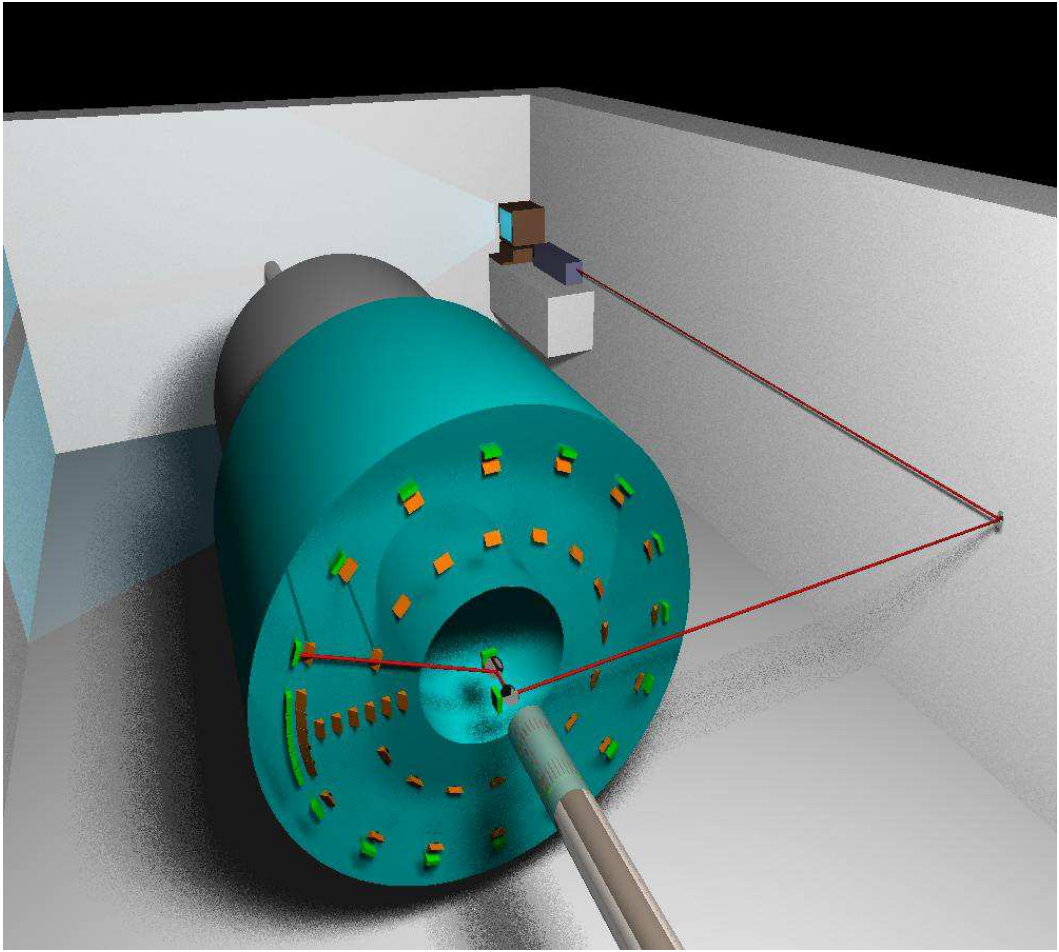}
\ecen
\caption{
Schematic view of the CERES experimental area as seen from downstream. 
The laser and the control PC are located in the far corner of the zone. 
From there the beam is transported to the distribution point where a 
rotating mirror sends the beam toward one of the azimuthal sectors of 
the TPC. Partially reflecting mirrors inject the beam into the TPC, 
parallel to the cylinder axis. 
The sector at about 9 o'clock, equipped with 11 mirrors, is referred to as 
''special sector''.}
\label{fig_zone-view}
\end{figure}

\begin{figure}[t]
\hspace*{-7mm}\scalebox{0.9}{\includegraphics{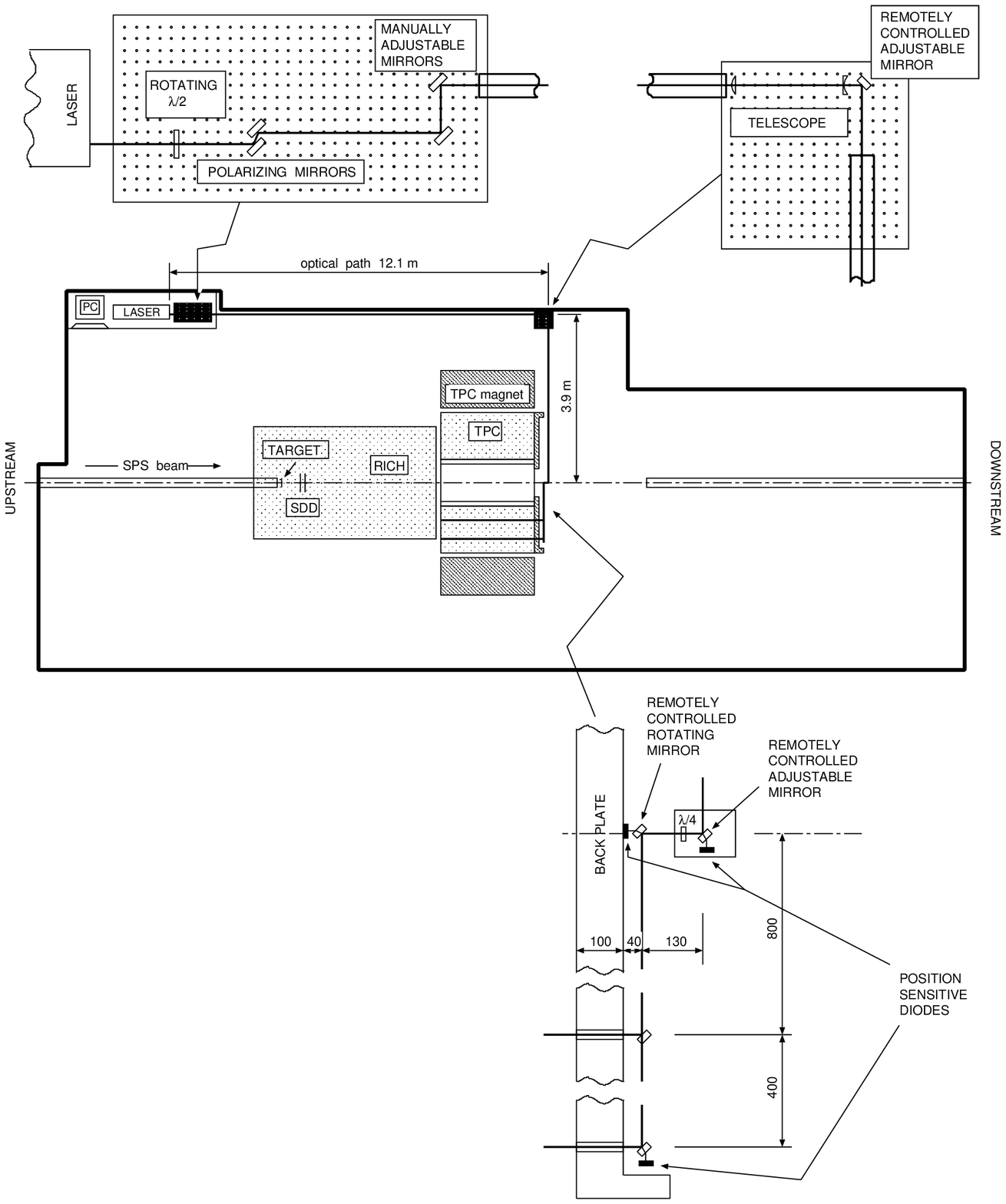}}\\
\caption{
Laser system layout. 
The laser beam travels 16~m before it arrives at the TPC. 
The mirrors are remotely controlled. 
Position sensitive diodes, placed behind the mirrors, monitor 
the beam position.}
\label{fig_zone}
\end{figure}

The upgraded CERES experimental setup is described in detail 
in~\cite{tpc-nim}.  
A pictorial representation of the experimental area, with emphasis on 
the components pertinent to the laser system, and a more detailed sketch 
of the setup are shown in Figs.~\ref{fig_zone-view} and \ref{fig_zone}, 
respectively. 
A laser generates short pulses of UV light. 
The laser beam is brought to the accelerator beam axis, which coincides with  
the symmetry axis of the TPC, at a location downstream of the TPC. 
From there the laser beam goes upstream along the symmetry axis 
up to the distribution point, located 4~cm downstream of the TPC backplate. 
At the distribution point the beam is reflected at an angle of 90\degr\ in 
$\phi$-direction determined by the orientation of a rotating mirror. 
The radial beam goes through one or more 45\degr\ partially reflecting 
mirrors which send the light into the TPC, parallel to its axis, 
through quartz windows. 
Fifteen azimuthal TPC sectors have two such mirrors each, giving two 
laser rays at the $r$= 800 and 1200~mm, both at the azimuthal center 
of the sector ($\phi=0$). 
The sixteenth sector (''special sector'') has 7 mirrors at $\phi=0$ at 
radii of 700, 800, 900, 1000, 1100, and 1200~mm, 
and 4 mirrors at $R=1200$~mm and $\phi=\pm$ 5\degr\ and $\pm$ 10\degr.
The laser beam is monitored using position sensitive diodes placed behind 
the mirrors. 
The individual components of the system are described in the following 
sections. 

\section{Laser}
\label{sec_laser}
\begin{figure}[h]
\hspace*{5mm}
\scalebox{0.65}{\includegraphics{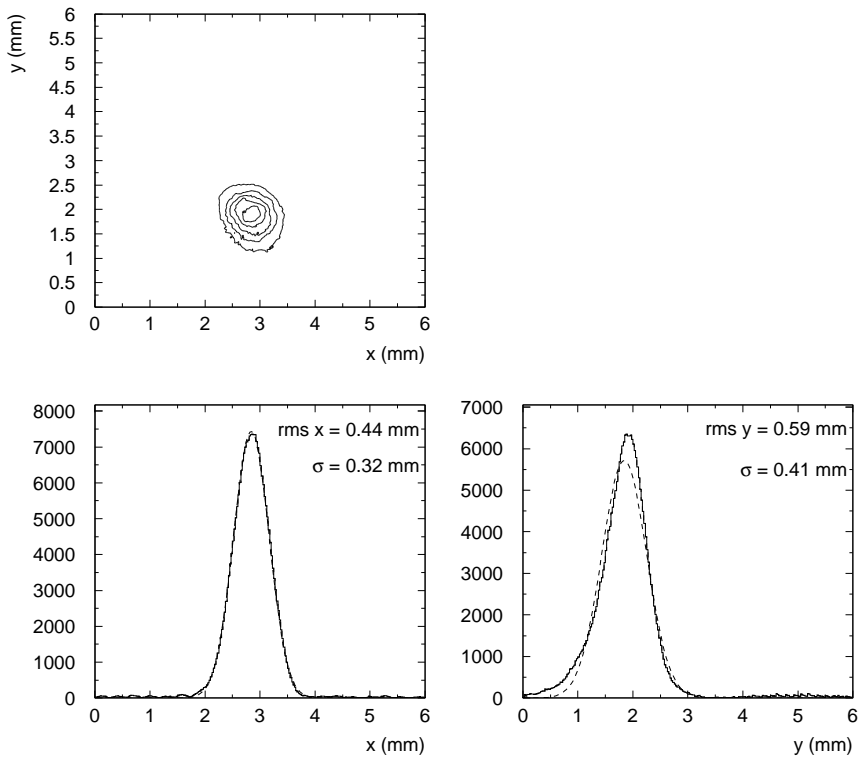}}
\caption{
The transverse intensity profile  (top) of the laser beam and its two 
projections (bottom). 
The standard deviations of the fitted Gaussian are $\sigma_x=0.32$~mm and 
$\sigma_y=0.41$~mm. 
The measurement was done at the exit from the laser. }
\label{fig_dm16}
\end{figure}

A Nd:YAG laser with two frequency doublers~\cite{laser} was used to generate 
4~ns long pulses of $\lambda$=266~nm light with an energy of up to 3-4~mJ per 
pulse and a repetition frequency of up to 10~Hz. 
The laser was chosen for its narrow beam and low divergence, essential 
for simulating particle tracks in a large drift chamber~\cite{na35}.  
A 2~mm diameter aperture located inside the laser cavity 
enforced the TEM$_{00}$ mode, resulting in a nearly-Gaussian beam profile. 
The nominal beam diameter $a$ (for Gaussian beams equal to 4$\sigma$) 
was 2~mm. 
The divergence, defined as the increase in diameter per unit 
path length in far field, was specified to be within 30\% of the 
diffraction limit, which, for Gaussian beams, is
\beq
\Theta_D = \frac{\d a}{\d z} = \frac{4 \lambda}{\pi a_0}.
\eeq
In the above formula $\Theta_D$ is the divergence angle, $a(z)$ is the 
beam diameter at position $z$, $\lambda$ is the wavelength, and $a_0$ is 
the beam waist (diameter of the beam at the exit of the laser). 
Diffraction thus limits the product of the waist diameter and the 
divergence of a Gaussian 266~nm beam to be 
$\Theta_D\cdot a_0=0.34$~mm$\cdot$mr. 
Indeed, depending on the adjustment one could produce diameters of 0.8-2~mm, 
larger beams having smaller divergence. 
Under typical conditions the diameter was $a_0=1.2$~mm and the 
divergence $\Theta_D=0.64$~mr, i.e. about twice the diffraction limit. 
A representative beam profile, measured with a laser camera~\cite{camera}, 
is shown in Fig.~\ref{fig_dm16}.

\begin{figure}[h]
\bcen
\scalebox{0.65}{\includegraphics{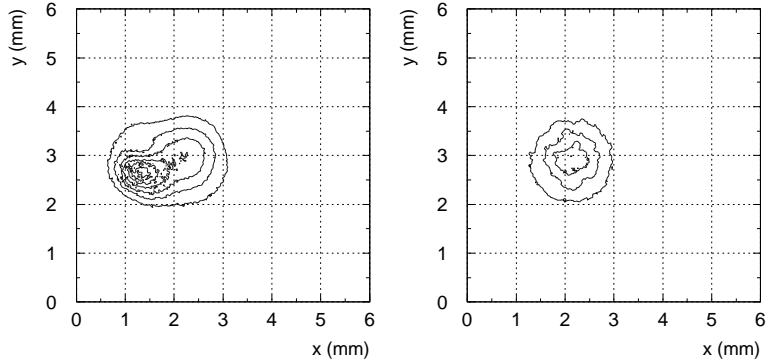}}
\ecen
\caption{
Left: laser beam intensity profile measured with (left) 
and without (right) an IR filter in front of the camera. 
The IR and the green beams are obviously not aligned.}
\label{fig_dm89}
\end{figure}

The primary laser beam was infrared (IR), $\lambda=1064$~nm. 
Two frequency doublers converted IR to green, $\lambda=532$~nm, 
and the green to UV, $\lambda=266$~nm. 
The IR, the green, and the UV beams had a linear 45\degr, a vertical, and 
a horizontal polarization, respectively. 
Since frequency doublers convert only a part of the incident light, 
a prism, installed inside the laser chassis after the second doubler, 
was employed to separate the UV light from the other two wavelengths. 
Suppressing the unwanted wavelengths is essential since the three 
components are, in general, not aligned (Fig.~\ref{fig_dm89}) 
and the standard optics is usually more sensitive to green than to 
UV\footnote{
The two dichroitic mirrors employed first for this purpose were still 
leaving about 80~$\mu$J of IR and 30~$\mu$J of the green light in the 
beam (compared to 4 mJ UV) leading to problems when measuring the beam 
energy or profile.}.

\section{Beam intensity adjustment}
\label{sec_attenuation}
\begin{figure}[h]
\vspace{5mm}
\bcen
\scalebox{1.1}{\includegraphics{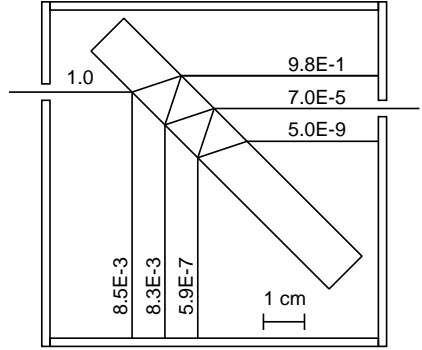}}
\ecen
\caption{
Multiple reflection attenuator. 
The laser beam enters from the left. The attenuated beam exits through 
a hole in the wall. The attenuation factors were calculated using the 
Fresnel equations with the refraction index of quartz n(266~nm)=1.4997 
and a 45\degr\ incidence angle. 
}
\label{fig_attenuator}
\end{figure}

For many applications the laser beam needs to be attenuated. 
For this purpose we built a simple device 
based on multiple reflections in a quartz plate surrounded by black walls 
except for the entrance and the exit holes (Fig.~\ref{fig_attenuator}). 
The attenuation factor was selected by choosing a wall with the appropriate 
exit hole location.  

\begin{figure}[h]
\bcen
\scalebox{0.6}{\includegraphics{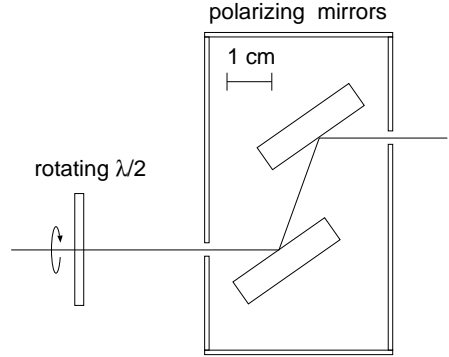}}
\ecen
\caption{
Variable attenuator (top view). 
The laser beam goes through a $\lambda$/2 plate which, depending on its 
azimuthal orientation, can rotate the polarization plane of the beam. 
The two mirrors reflect only the vertical polarization component. 
This way the intensity of the outgoing beam depends on the orientation 
of the $\lambda$/2 plate.}
\label{fig_polarizer}
\end{figure}

\begin{figure}[h]
\hspace*{5cm}\scalebox{0.65}{\includegraphics{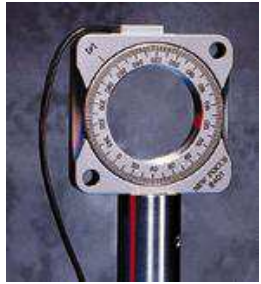}}
\caption{
Remotely controlled rotary stage~\cite{rotary_stage}. The driving mechanism 
is based on the piezoelectric effect. }
\label{fig_rotary_stage}
\end{figure}

The fixed and large attenuation factors obtained using multiple 
reflections were of advantage during the development of the laser 
system components when the absolute calibration was needed. 
For the final setup in the experimental area a simple attenuator 
consisting of a rotating half-wave plate and two polarizing mirrors 
(Fig.~\ref{fig_polarizer}) was used to adjust the intensity based on the 
TPC response. 
The half-wave plate rotated the polarization plane of the passing beam from 
its original horizontal orientation by an angle which was related to the 
azimuthal orientation of the half-wave plate itself. 
The plate was mounted in a rotary stage (Fig.~\ref{fig_rotary_stage}). 
The rotary stage, and with it the polarization plane of the transmitted 
laser beam, could be adjusted remotely. 
The mirrors were high reflectivity (HR) coated for S-polarization, 
and the reflection occurred at the Brewster angle such that only the 
vertically polarized light survived the two reflections. 
The half-wave plate/mirrors combination thus provided a remotely controlled 
variable attenuator with the attenuation factor between 0 and 1. 

Note that the laser itself provided two possibilities of adjusting 
the intensity: by changing the charger voltage and by changing the 
Q-switch delay. The first affected the flash lamp intensity. 
The latter changes the time elapsed between the flash and the laser action. 
During this time the spontaneous emission would lower the number of excited 
molecules and thus the longer one waited the less energy remained for 
the laser pulse. 
However, too long and too short delays resulted in high pulse-by-pulse 
intensity fluctuations and bad beam quality, respectively. 
In addition, the delay value affected the relative timing between the light 
pulse and the electric signal generated by the Q-switch electronics 
which was used to trigger the experiment data acquisition. 
For this reason we kept the delay value within 25-50\% of its maximum, and 
adjusted the intensity by the two devices described above. 

\section{Beam transport from the laser to the axis of the TPC}
\begin{figure}[h]
\bcen
\scalebox{0.65}{\includegraphics{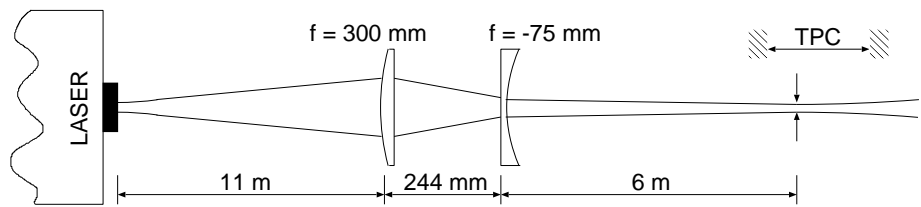}}
\ecen
\caption{
Beam focusing (not to scale). 
At the Galilean telescope, located 11~m after the laser, 
the beam has a diameter of about 1~cm. 
The telescope focuses it back to, approximately, the initial diameter 
($\sim$1~mm) inside the TPC.}
\label{fig_telescope}
\end{figure}

After the polarizers the laser beam was reflected by two manually adjustable 
45\degr\ mirrors. 
Subsequently, the beam was led out of the laser tent, located in a corner 
of the zone, 
and went about 11~m along the wall, parallel to the TPC symmetry axis 
and at the same height as the particle beam axis 
(see Figs.~\ref{fig_zone-view},~\ref{fig_zone}). 
At this point the beam reached a platform, 
attached to the wall, on which two lenses and a mirror were mounted. 
The lenses, which formed a Galilean telescope, were set up such that the 
beam was focused approximately 6~m from there, i.e. within the 
downstream half of the TPC. 
This location of the focus was chosen to keep the beam diameter small 
at the position sensitive diodes which monitor the beam position before 
the entrance to the TPC (see the next two sections).
The transverse beam profile is shown in Fig.~\ref{fig_telescope}. 

\begin{figure}[h]
\bcen
\scalebox{0.65}{\includegraphics{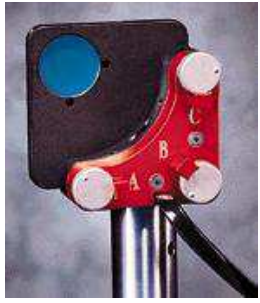}}
\ecen
\caption{Remotely controlled mirror holder~\cite{mirror_holder}. 
The three actuators can be rotated manually or by applying series of 
asymmetric pulses which are converted to mechanical motion via the 
piezoelectric effect. }
\label{fig_mirror_holder}
\end{figure}

After passing the telescope, the beam was reflected by a 45\degr\ mirror 
mounted in a remotely controlled mirror holder (Fig.~\ref{fig_mirror_holder}), 
and sent toward the symmetry axis of the TPC. 

\section{Distribution point}
\label{sec_distribution_point}
\begin{figure}[h]
\bcen
\scalebox{1.5}{\includegraphics{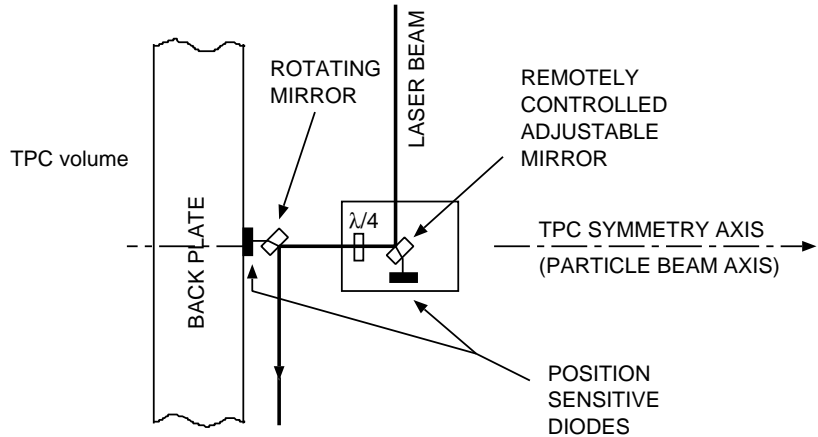}}
\ecen
\caption{Optics at the distribution point, top view.}
\label{fig_zone-zoom}
\end{figure}

The optics mounted on the TPC symmetry axis is shown in 
Fig.~\ref{fig_zone-zoom}. 
The beam, which arrives horizontally from the wall of the experimental area,   
hits a partially reflecting mirror 
located on the TPC symmetry axis 17~cm downstream of the backplate 
(see Figs.~\ref{fig_zone-view},~\ref{fig_zone}). 
The mirror is mounted in a remotely controlled holder of the type shown 
in Fig.~\ref{fig_mirror_holder}. 
The holders and the rotary stages used are driven by the piezoelectric effect 
and thus are not affected by a magnetic field. 
The transmitted beam hits a position sensitive diode. 
The reflected beam passes through a quarter-wave plate which transforms 
the linear polarization into a circular one. 
The advantage of the circular polarization is that the coefficients of the 
subsequent reflections are independent of the orientation of the rotating 
mirror. 
The rotating mirror reflects the beam radially toward one of the TPC sectors. 
The azimuthal angle of the mirror, and with it the sector of the TPC, 
can be selected remotely. 
Again, the mirror is semi-transparent, with a position sensitive diode 
placed behind it. 
The reflected beam is parallel to the backplate, 4~cm from its surface. 

\begin{figure}[h]
\bcen
\hspace{-5mm}\includegraphics[width=5cm]{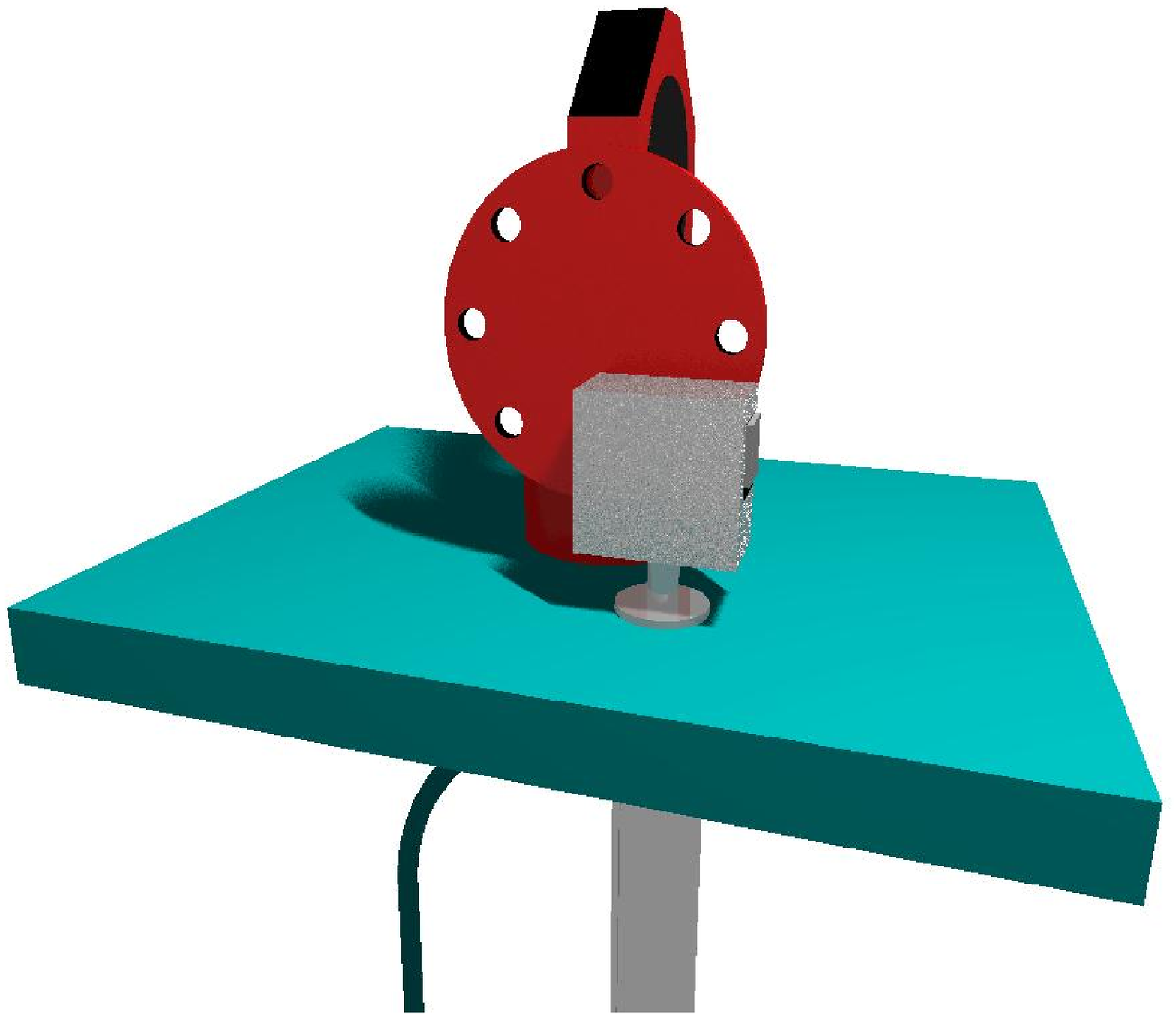}
\hspace{-5mm}\includegraphics[width=5cm]{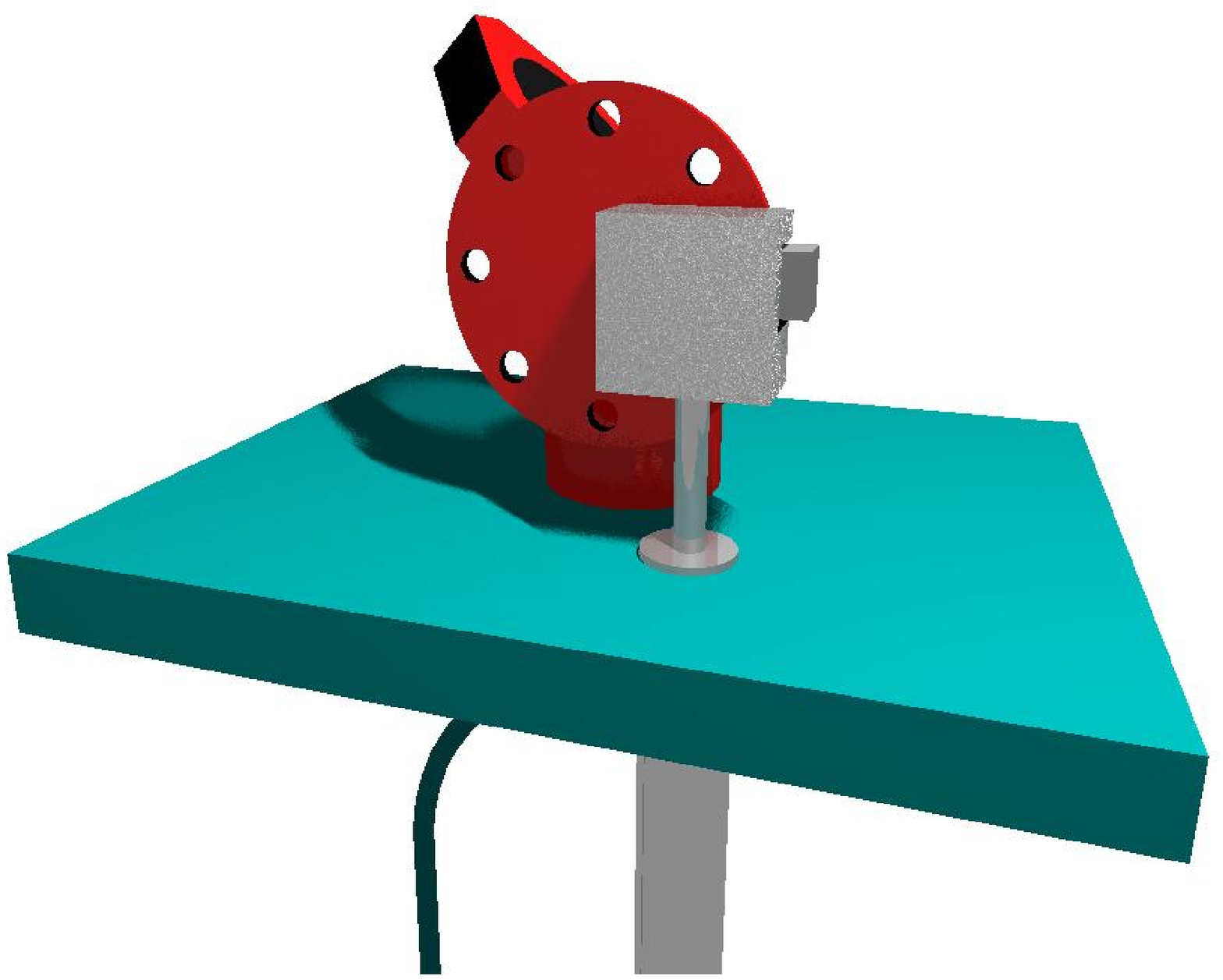}
\hspace{-5mm}\includegraphics[width=5cm]{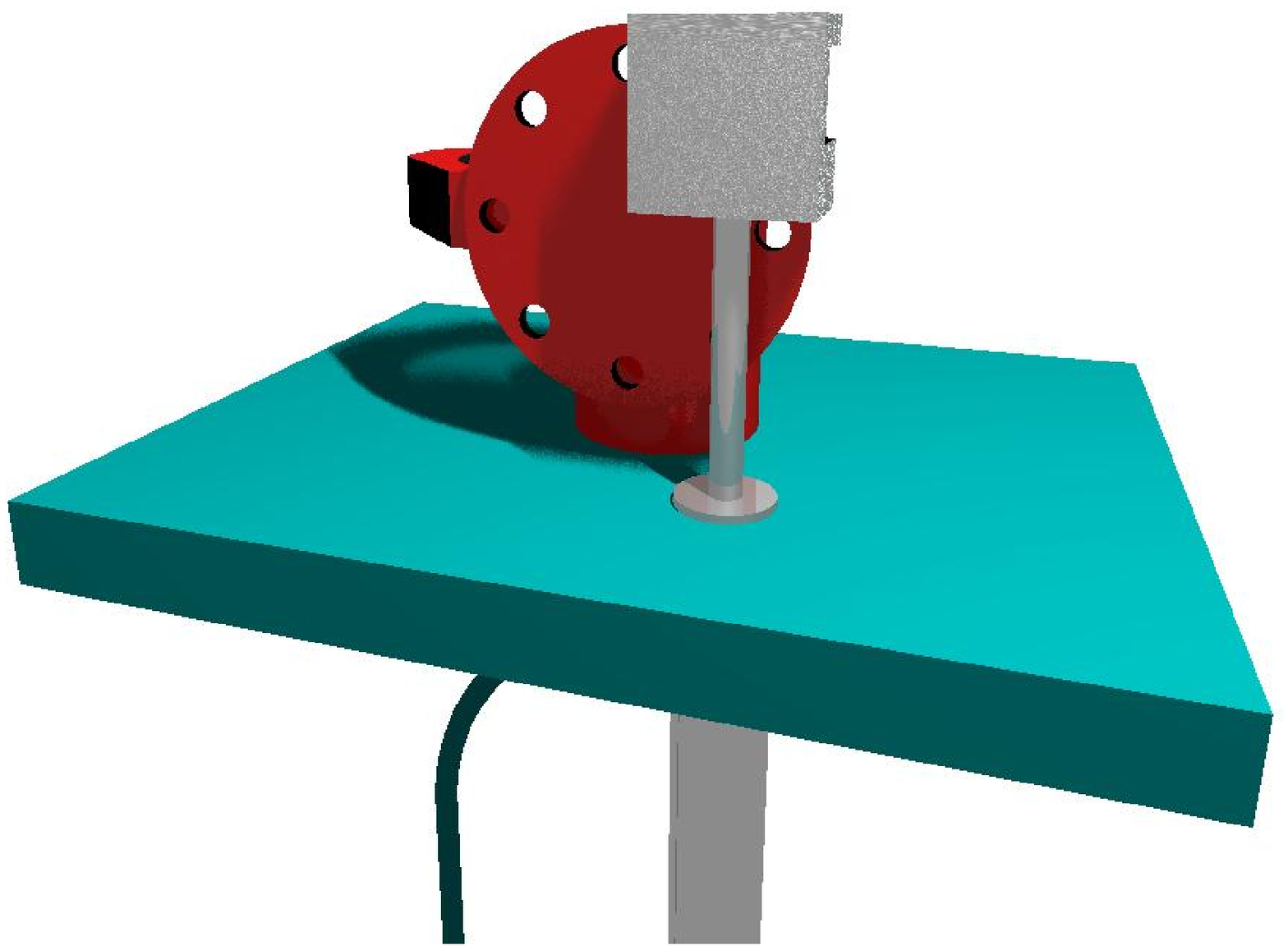}
\ecen
\caption{Manual flipper~\cite{flipper} driven by a pneumatic actuator.
For computer animation see \cite{flipper-movie}. }
\label{fig_flipper}
\end{figure}

\begin{figure}[h]
\hspace*{-3mm}
\scalebox{1.1}{\includegraphics{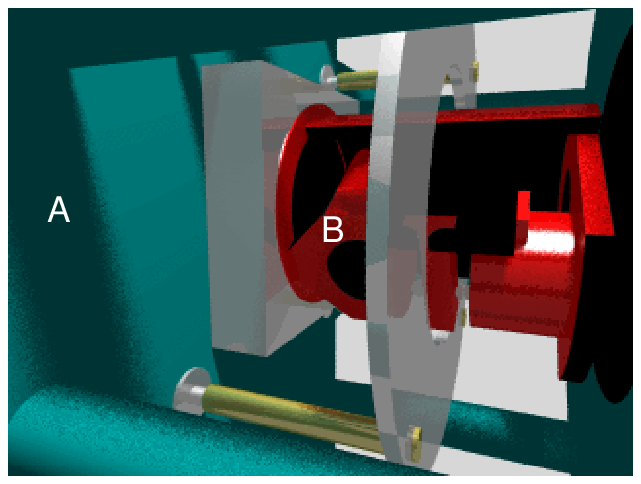}}
\scalebox{1.1}{\includegraphics{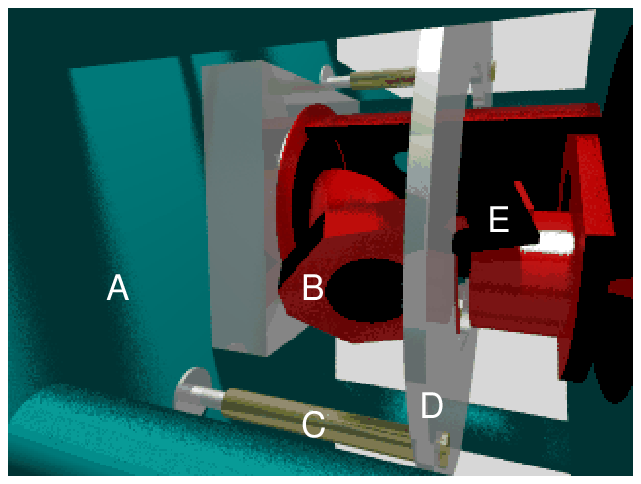}}
\caption{Rotating mirror at the distribution point. 
The laser beam comes from the right and is reflected by 90\degr\ 
in an azimuthal direction defined by the orientation of the rotating 
mirror holder B, in this case (left panel) into the plane of the figure 
and somewhat up. 
For the duration of the ion beam spill the mirror is removed from the axis. 
For this, the four rods (C) driven by pneumatic actuators move the 
aluminum ring (D)
which pushes the hook (E) of the rotating mirror holder such that the 
holder tilts and moves the mirror out of the beam axis, independently 
of its azimuthal orientation. 
For computer animations see \cite{otron-movie}. }
\label{fig_clutch}
\end{figure}

The design, which made use of the cylindrical symmetry of the TPC, 
thus required that two mirrors, a diode, and a quarter-wave 
plate stay on the particle beam axis. 
Monte Carlo simulations predicted a non-negligible background  
in the TPC resulting from the nuclear reactions in these elements 
unless they were kept out during the SPS spill. 
Since motorized flipping mirror holders like~\cite{motorized_flipper} 
would not work because of the magnetic field 
we used pneumatic actuators and remotely controlled valves 
to move non-motorized flippers, as shown in 
Fig.~\ref{fig_flipper}. 
Three of the four optical elements located in the beam axis were operated 
as shown in this figure. 
The rotating mirror, however, had to be movable independently of its 
azimuthal orientation. 
The solution is shown in Fig.~\ref{fig_clutch}. 
Four pneumatic actuators were used to move an aluminum ring 
which pushed a hook of the central mirror holder, tilting the holder 
and moving the mirror out of the beam axis. 
In all cases the mechanics has been designed such that the final mirror 
positions were defined by the end-points of the holders and not by the 
actuators. 

The pneumatic actuators were of the double acting type~\cite{festo-cylinder}. 
Two plastic tubes with an internal diameter of 2.9 mm supplied air to each 
of them. 
The air flow was regulated by PC-controlled valves. 
The air pressure was 2~bar and the air flow was reduced at the 
entrance to the cylinders for gentle operation. 

The system was designed such that the mirrors would be moved into the 
beam at the beginning of the spill pause, 
the central mirror would turn toward the next TPC sector, 
one laser pulse would be sent with simultaneous generation of the TPC trigger, 
and finally the mirrors would be moved out in time for the next spill. 
This indeed worked at the maximum SPS energy where the spill pause 
was about 14~s. 
At 40\agev\ the spill pause was just long enough to move the mirrors in and 
out but not to rotate the mirror, and thus the laser events had tracks 
generated in just one sector of the TPC. 
Dedicated laser runs turned out anyway to be more useful than running in 
parallel with physics data-taking. 

Optical and acoustical inspection of the distribution point was possible 
by means of an ordinary black-and-white camera~\cite{conrad-camera}, 
mounted about 30~cm from the axis and connected to a TV-set in the 
experiment's control room. 

\section{Entrance to the TPC}
\label{sec_entrance}
\begin{figure}[h]
\bcen
\scalebox{0.9}{\includegraphics{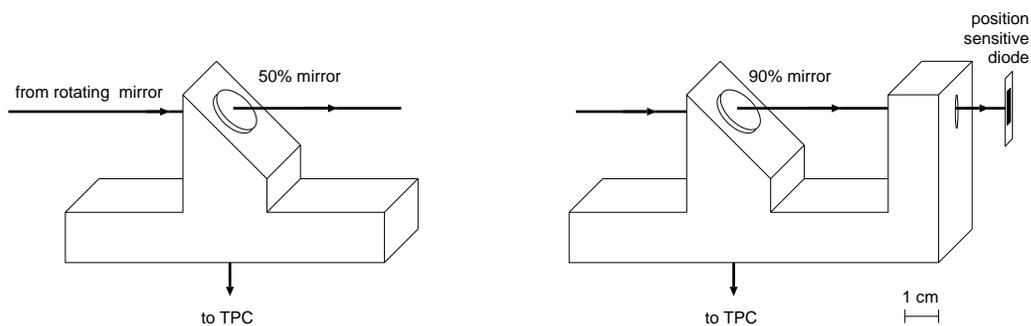}}
\ecen
\caption{Fixed mirrors which send the laser rays into TPC volume. 
The mirrors at $r$ = 800 mm (left panel) reflect 50\% of the light 
into the TPC. The mirror at $r$=1200 mm (right panel) reflects 90\% 
of the remaining light into the TPC, the rest hits the position 
sensitive diode.} 
\label{fig_holders}
\end{figure}

\begin{figure}[h]
\hspace{5mm}
\scalebox{1}{\includegraphics{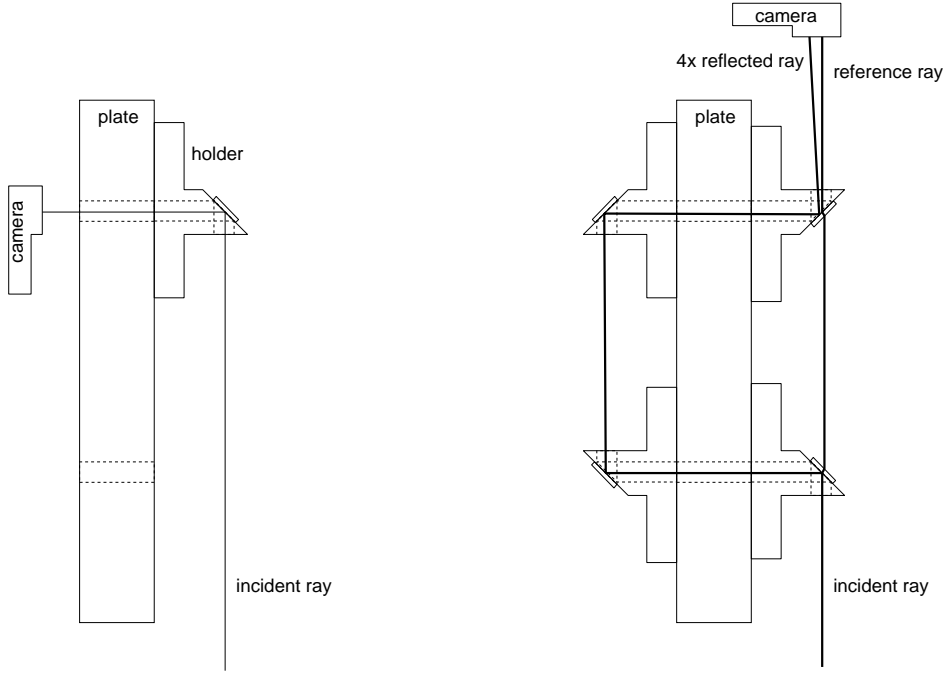}}
\caption{Relative (left) and absolute (right) mirror angle calibration 
in the laboratory. 
The angles of the reflected rays were measured by placing the camera at 
two different distances. }
\label{fig_holder-calibration}
\end{figure}

The laser beam, reflected from the rotating mirror at the TPC axis, 
passed through partially reflecting mirrors which sent 
the rays into the TPC (Fig.~\ref{fig_holders}). 
These mirrors were glued to massive aluminum holders, mounted on the 
TPC backplate with bolts. 
Stable geometry was essential because the laser ray position was 
monitored only before its entrance into the TPC, and the position inside 
was calculated using the calibrated mirror position and angle. 
The holders were calibrated in the laboratory by measuring the positions 
and the angles of the reflected rays as shown in 
Fig.~\ref{fig_holder-calibration}. 
The reflected ray angles were measured via ray position at two different 
distances.
The two reflection angles were first determined for each mirror holder 
relative to a reference holder. 
Subsequently, the holders were mounted in groups of four on a massive plate 
and the sum of the four reflection angles was measured absolutely. 
From the four relative and the one absolute measurement absolute 
reflection angles were determined for all holders. 
To cancel the related mounting uncertainties 
the measurements were repeated with different orientations of the 
support plate. 

The TPC backplate was made of cast aluminum and had a thickness of 10~cm. 
The mirror holders were mounted on it with precision bolts. 
The exact positions of the mounting holes were measured before assembling 
the TPC. 
For this, a carousel-like custom tool and large calipers were used to 
determine the center of the backplate and to measure the radial positions 
of the mounting holes. Furthermore, mutual distances between the holes 
were determined. 
The 151 individual measurements were fit by a geometrical model with 93 
free parameters, corresponding to the 90 radial $r$ and azimuthal $\phi$ 
coordinates of the 45 holes, 
4 unknown measurement constants, and arbitrarily fixing the $\phi$ 
coordinate of one hole. 
The chi-square of the fit allowed to estimate the precision of the 
obtained hole positions $dr$ and $r d\phi$ to be 0.1~mm and 0.2~mm, 
respectively. 
Deviations from the design position as large as 0.5~mm, i.e. ten times 
exceeding the specified limit, were observed. 

As the next step, the mirror holders were mounted and the laser was turned on. 
The ray positions on the downstream side of the backplate were 
fixed using collimators. 
The rays emerging from the backplate on the upstream (TPC) side were measured 
at two distances, 3~cm and 30~cm. 
The obtained angles showed a clear correlation with the laboratory 
measurement of the mirror angles. 
Based on the width of the correlation and on the hole position uncertainties, 
the resolution, with which the absolute ray position inside the TPC can be 
calculated for a known ray position outside, 
was estimated to be 0.25~mm and 1~mm 
at the backplate and at the upstream end of the TPC, respectively. 

The laser rays traversed the active volume of the TPC upstream 
parallel to the symmetry axis. 
At the end of their trajectory the rays would hit the upstream field cage 
foil. 
As was found in laboratory tests, 100 laser pulses with an energy 
of 260~$\mu$J and a beam diameter of $a$=2~mm were sufficient to produce a 
black spot on the kapton foil. 
With full energy (4 mJ), 600 pulses were enough to burn a hole. 
In order to protect the field cage, beam dumps made of ordinary microscope 
cover glass with a thickness of 160 $\mu$m were glued to the upstream foil at 
the locations corresponding to the quartz windows in the downstream foil. 

\section{Monitoring of the laser beam position}
\label{sec_diodes}
The calibration of the TPC was based on the assumption that the absolute 
position of the laser rays inside the drift chamber was known. 
For this, the massive backplate of the TPC was used as a reference. 
The geometrical calibration of the backplate and of the mirrors mounted 
on it was described in Section~\ref{sec_entrance}. 
Here we describe the way of monitoring the position of the laser beam 
with respect to these fixed elements. 

\begin{figure}[h]
\bcen
\scalebox{0.8}{\includegraphics{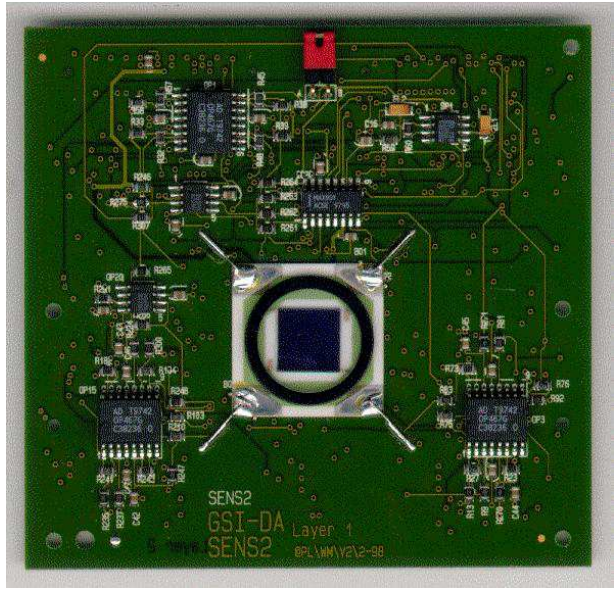}}
\ecen
\caption{Position sensitive diode~\cite{diodes} on the readout 
board~\cite{diode-electronics}.}
\label{fig_diode}
\end{figure}

The coating of the dielectric mirrors at the distribution point and 
of the TPC entrance mirrors at $r$=1200~mm was chosen such that 10\% of the 
light was transmitted. 
Windowless duo-lateral position sensitive diodes~\cite{diodes}, 
placed behind the mirrors, were used to monitor the laser beam. 
The active area of the diodes was 10$\times$10~mm$^2$ and the nominal position 
resolution was better than 1~$\mu$m. 
Hit by a laser beam, each diode produced two cathode and two anode signals. 
The sum of amplitudes was proportional to the light intensity, 
and the difference between the two cathode (anode) signals was related to 
the horizontal (vertical) position of incidence. 
The coordinates were reconstructed from the measured right, left, up, and 
down signals $r$, $l$, $u$, $d$ via  
\beq
    x&=5~{\rm mm} \cdot &(r-l)/(r+l) \nonumber\\
    y&=5~{\rm mm} \cdot &(u-d)/(u+d) . 
\eeq
The signals were amplified, stretched, and digitized by electronics 
located on two 10$\times$10~cm$^2$ large boards~\cite{diode-electronics}. 
The diode itself was mounted on the first board (Fig.~\ref{fig_diode}). 
Precision bolts ensured exact positioning of the mounted boards. 
The boards were connected via a daisy chain to a PC. 
The readout was first triggered internally by the enabled diodes. 
This was later replaced by an external trigger derived from the laser 
Q-switch. 
After each event the data of all diodes were transported to the memory of 
the PC via an interface connected to the parallel port. 

\begin{figure}[h]
\hspace*{1cm}\scalebox{1}{\includegraphics{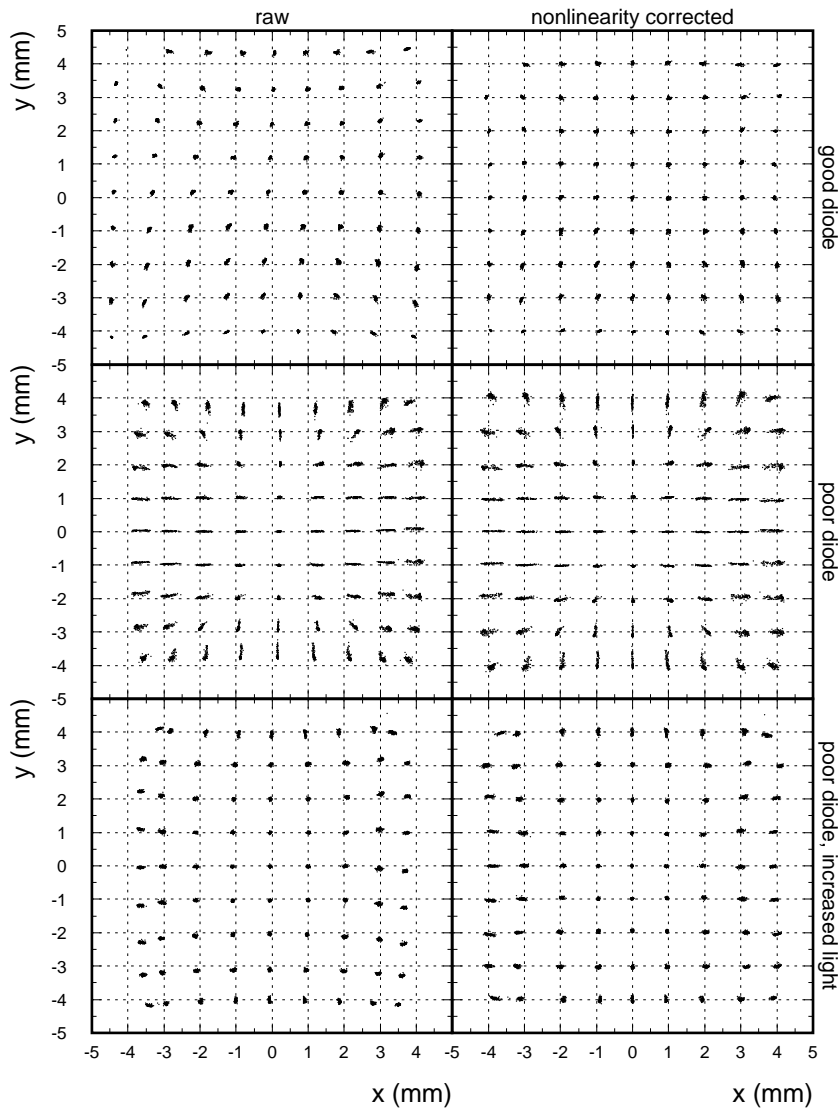}}
\caption{Laser beam position recorded during a scan of the position 
sensitive diodes, raw (left) and after calibration (right). 
The uppermost row shows a good diode from the first production batch. 
Diodes from the second batch showed a much worse resolution (middle row) 
but recovered when illuminated with a higher light intensity (bottom row). }
\label{fig_diosca}
\end{figure}

The linearity and the resolution of the diodes were tested in laboratory 
by illuminating each diode at 81 positions on a 1~mm grid 
(Fig.~\ref{fig_diosca}). 
For this, each diode in turn was mounted on a pair of computer-controlled 
linear stages. 
The calibration procedure was fully automatic and consisted in shifting 
the diode to each of the 81 positions, switching on the laser, and 
recording 100 diode events. 
The typical resolution of a single spot was 
$\sigma_x\approx\sigma_y\approx30~\mu$m. 
The global resolution after the calibration was about 50~$\mu$m. 
The laser beam energy during such a measurement was typically 1~$\mu$J. 

The second batch of the diodes delivered turned out to require as much 
as 10~$\mu$J in order to achieve a good resolution. 
These diodes were assigned to the locations at which the intensity  
of the incident laser beam was higher. 

\begin{figure}[h]
\hspace*{3cm}\scalebox{0.7}{\includegraphics{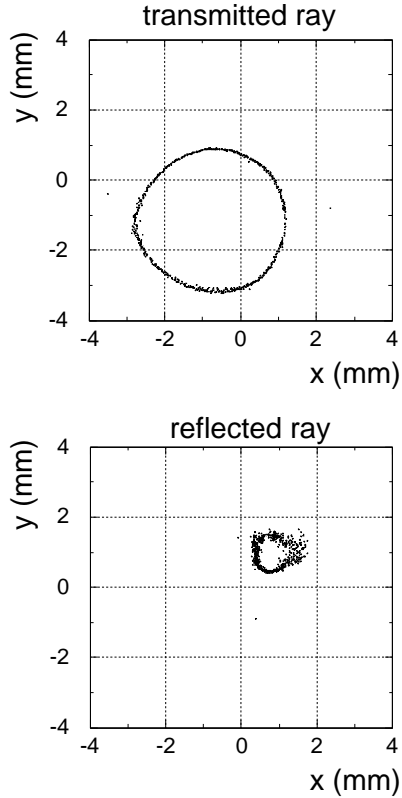}}
\caption{Position of the laser beam transmitted through (up) and 
reflected by (bottom) the rotating mirror. The finite radius of the 
upper circle results from the finite thickness of the mirror. 
The lower circle shows that the laser beam is not on the rotation 
axis. See text for details.}
\label{fig_gotoaxis}
\end{figure}

One of the diodes was placed on the rotating mirror holder, 
with the connections to the readout electronics provided via 
four sliding pins, 
and used to monitor the position of the reflected beam. 
A dedicated semi-transparent mirror, placed on the laser beam right after 
the rotating mirror, was reflecting a small fraction of the laser light 
onto this diode. 
(For simplicity these two elements were left out in Fig.~\ref{fig_zone-zoom}.)
During a rotation of the mirror the spot on this diode would follow a circle  
whose radius would indicate the deviation of the laser beam from 
the mirror rotation axis and thus from the symmetry axis of the TPC. 
An example is shown in the bottom panel of Fig.~\ref{fig_gotoaxis}. 
The upper panel of the same figure shows the position of the transmitted ray, 
displaced after having traversed the 6~mm thick mirror at an incidence angle 
of 45\degr. 
The information from this diode was used to monitor the mirror angle 
during the rotation. 
The correction necessary to bring the beam to the axis could be 
calculated from the data collected by these two diodes during a 2$\pi$ 
rotation, combined with the $z$-coordinates of the reflected 
beam measured on the diodes distributed on the circumference of the TPC. 
This procedure could be applied iteratively; in praxis, one iteration 
was sufficient to place the incident beam exactly (better than 100~$\mu$m)  
on the TPC symmetry axis. 

After the incident beam was placed on the axis, the position information 
from all diodes, combined with the known reflection angle of the mirror 
at the entrance to the TPC, 
was used to calculate the laser beam position in the TPC. 
This, in principle, could be done separately for each single laser event. 
However, with the excellent laser pointing stability (standard deviation 
at the entrance to the TPC of 60~$\mu$m) within the duration of a typical 
laser run (10 minutes), it was more convenient to use the average 
beam position of a run. 
This information,  combined with the knowledge of the mirror geometry, was 
used to determine the laser track positions in the chamber. 

\section{Computer control}
\begin{figure}[h]
\hspace*{3cm}\scalebox{0.5}{\includegraphics{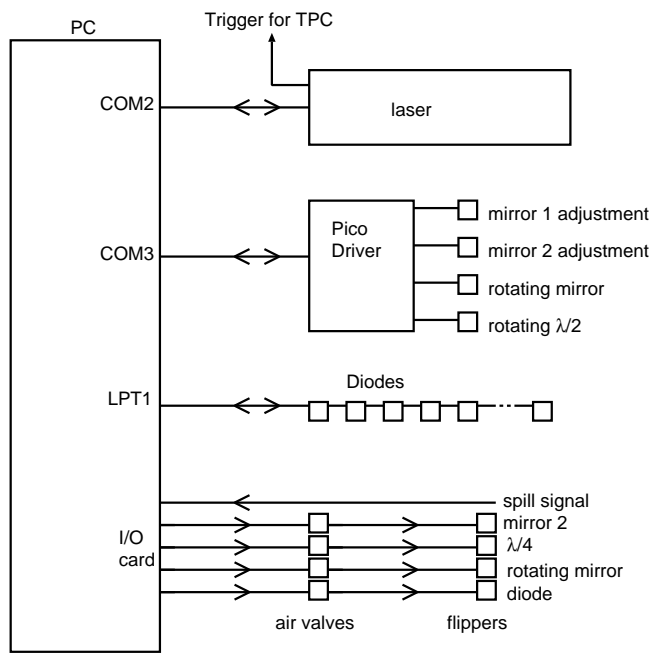}}
\caption{Control computer connection scheme.}
\label{fig_pc}
\end{figure}

A personal computer running under Linux was used to control the 
laser and the optomechanics, and to collect the diode data. 
Synchronization with the accelerator spill signal was achieved via 
an I/O card. 
The same I/O card was used to control the air valves of the 
actuators for moving mirrors in and out of the beam axis, described 
in Section \ref{sec_distribution_point}.
The logical scheme is shown in Fig.~\ref{fig_pc}. 

In laboratory tests the PC was used to read the CCD camera 
and to perform fully automatic diode scans involving moving linear 
stages, switching the laser, and collecting diode data.

\section{Trigger timing}
While in nuclear collision events the data acquisition trigger was 
generated using the incident beam particle, 
for laser events the trigger was derived from the Q-switch of the laser. 
The timings of these two triggers were not identical. 
The time difference was determined in two ways. 
First, during a dedicated test a 0.5 mm diameter light guide was used 
to connect the laser with a scintillator located some 5 m downstream 
of the TPC such that the attached phototube was seeing a scintillation 
signal in collision events and laser light in laser events. 
After correcting for the time-of-flight of particles and for the light 
propagation velocity in the light guide, in a separate test determined to be 
19.4~cm/ns, 
the time difference between the two triggers was calculated to be 
between 300 and 410~ns, depending on the laser settings, to be subtracted 
from the TPC drift times in laser events before they can be compared to 
nuclear collision events. 
The estimated accuracy of this measurement was 20~ns. 
The result was in good agreement with the number obtained by the second 
method which consisted in comparing the position of the peak caused by 
photoelectrons from the high voltage cylinder limiting the drift 
volume at low radii (see Section \ref{sec_first_tests}), seen in laser 
events, with the edge of the radial distribution of hits in collision events. 

\section{TPC commissioning with laser}
\label{sec_first_tests}
\begin{figure}[h]
\hspace*{5mm}\scalebox{0.8}{\includegraphics[angle=-90,clip=]{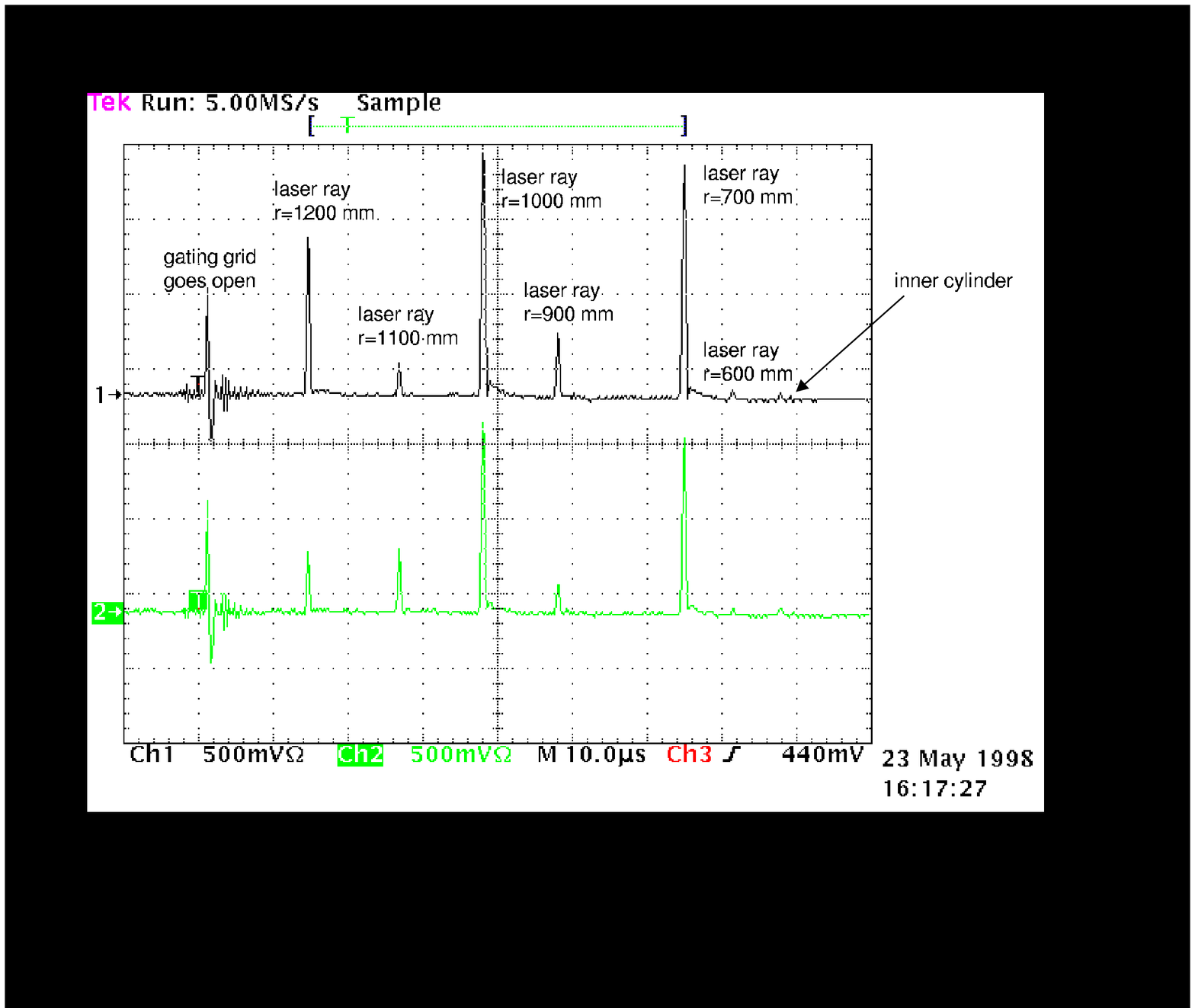}}\\
\caption{
Charge on two readout pads versus drift time. 
The bipolar pulse is due to switching of the gating grid.  
The peaks are caused by laser rays. Some of the laser rays are not 
properly centered and thus have a lower intensity. 
The ray at $r$=800~mm is not visible at all. 
The tiny peak on the right hand side (longest drift time) represents 
photoelectrons produced on the high voltage cylinder by the scattered 
laser light. }
\label{fig_las2}
\end{figure}

First laser tests of the TPC were performed in May 1998. 
The laser beam was sent into the special sector of the TPC. 
Every laser pulse produced seven parallel tracks at the same $\phi$ 
and at different radii. 
The response of two pads of the readout chamber is shown in 
Fig.~\ref{fig_las2}. 

\begin{figure}[h!]
\hspace*{2cm}
\includegraphics[width=10cm, height=6cm]{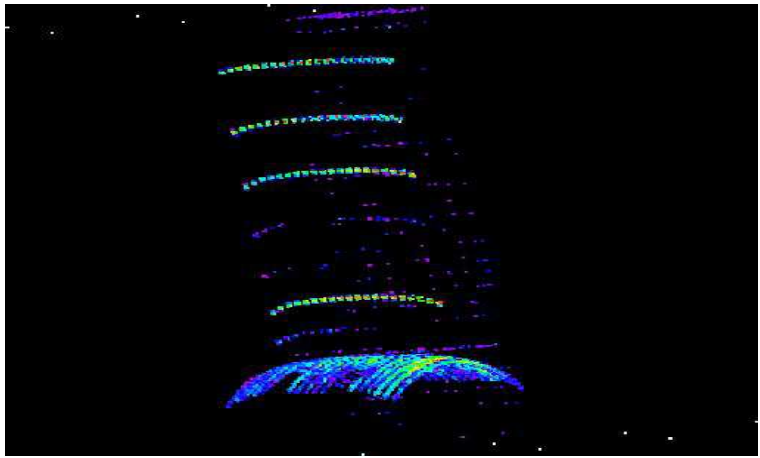}\\
\caption{
Laser tracks reconstructed in the first tests in the special sector of 
the TPC. }
\label{fig_laser329}
\end{figure}

In summer and fall of 1998 the TPC was fully equipped with readout electronics 
and various calibrations were performed with tagged muons and the laser. 
A typical laser event from that time is shown in Fig.~\ref{fig_laser329}. 
In addition to the tracks made by laser rays ionizing TPC gas the signal 
from the high voltage cylinder is visible, generated by photoelectrons from 
the aluminum (work function 4.28~eV) produced by scattered laser photons 
($\hbar\omega$=4.65~eV). 

Close to the TPC ends the reconstructed tracks exhibit curvature indicating 
a $z$-dependence of the electric field not accounted for by the 
reconstruction software. 
This observation triggered a new three-dimensional calculation of the 
electric field~\cite{tpc-nim}. 


\section{Pulse shape and grid transparency studies}
\label{sec_pulse_shape}
\begin{figure}[t!]
\hspace*{1cm}\scalebox{0.5}{\includegraphics{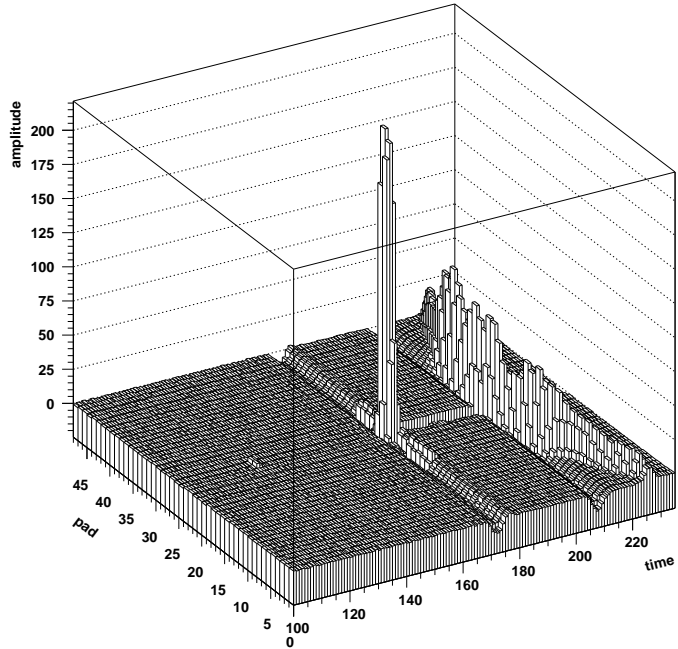}}\\
\caption{
Response of a readout chamber averaged over many laser events. 
The laser pulse is accompanied by an undershoot and the lateral crosstalk. 
The signal at high drift times, distributed over many pads, comes 
from photoelectrons knocked out of the high voltage cylinder by the 
scattered laser photons.}
\label{fig_schmitz5.3}
\end{figure}

The laser system allows to generate many events with a track at the same
position and with similar signal amplitudes. 
By averaging over many events in the absence of other tracks 
details of the pulse shape of the readout chamber can be studied. 
Two effects of interest, an undershoot following each pulse and 
a negative signal on pads coupling to the same anode wire 
(``lateral cross-talk'')
are shown in Fig.~\ref{fig_schmitz5.3} (taken from~\cite{schmitz}).
Also visible is the signal coming from electrons knocked out of the 
high voltage cylinder by stray laser light. 
The shape of the undershoot has been parametrized and corrected for. 
The possibility of accurately adjusting the laser beam intensity proved 
helpful in studying these subtle effects. 

Ions produced by electrons via gas amplification in the vicinity of the 
anode wires were kept out of the TPC drift volume by means of a gating 
grid which was opened only for the duration of the drift time (70~$\mu$s) 
after each  trigger. The wires of the gating grid had an offset potential 
of -140~V in the open state, matched to the electric field of the drift 
volume, and alternating potentials of -70 and -210~V (bias of $\pm$70~V) 
in the closed state. 
The voltages were chosen based on a measurement of the grid transparency 
in laser events. 
The procedure is described in detail in~\cite{tpc-nim}.  

\section{Calibration of the chamber position and of the electronics 
time offset}
\begin{figure}[b]
\hspace*{5mm}\includegraphics[width=13cm]{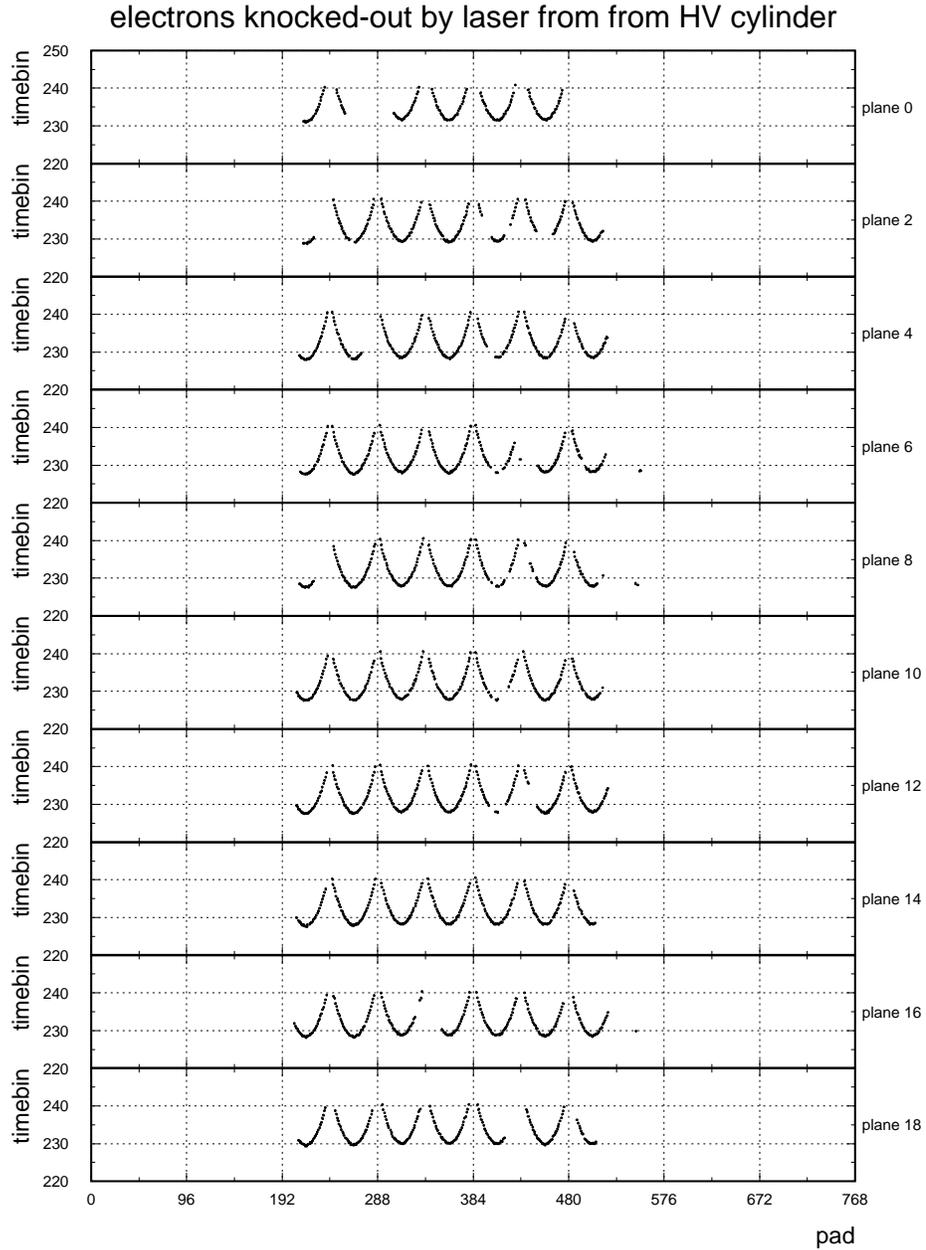}\\
\caption{The drift times of electrons emitted from the high voltage cylinder 
to individual pads (related to $\phi$) for several readout planes ($z$). 
The periodic structure is caused by the polygonal shape of the readout. 
The laser illuminates the high voltage cylinder only from one side. }
\label{fig_ana-cylinder-c}
\end{figure}

\begin{figure}[b]
\hspace*{5mm}\includegraphics[width=13cm]{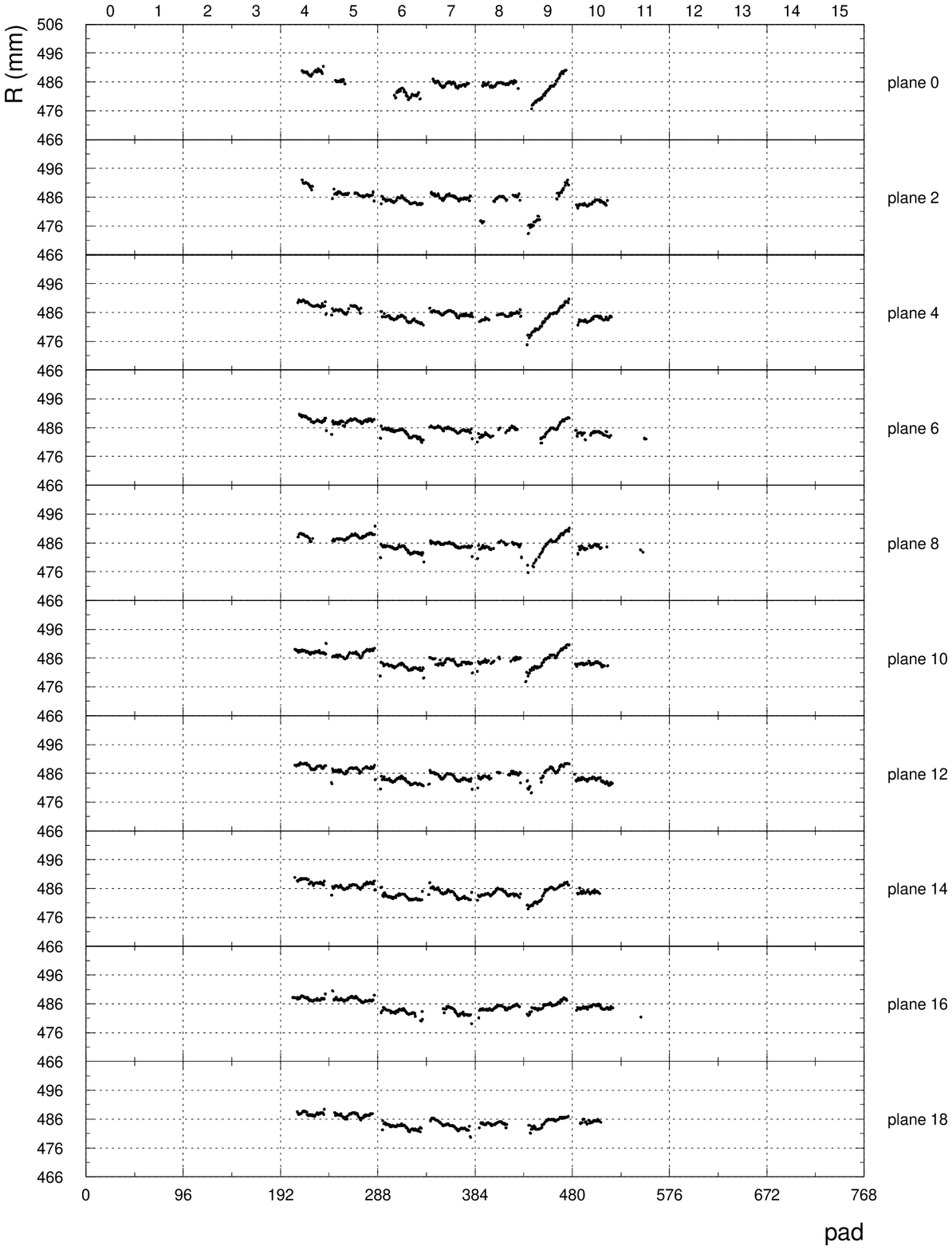}\\
\caption{
Radius of the high voltage cylinder reconstructed from the drift time of the 
photoelectrons (comp. Fig.~\ref{fig_ana-cylinder-c}). 
The numbers at the top indicate the sixteen readout chambers, 48 pads each. 
The structures caused by misaligned readout chambers and a periodic 
distortion within each chamber, related to three front-end-boards, 
are visible. }
\label{fig_ana-cylinder-e-raw}
\end{figure}

\begin{figure}[t]
\hspace*{5mm}\includegraphics[width=13cm]{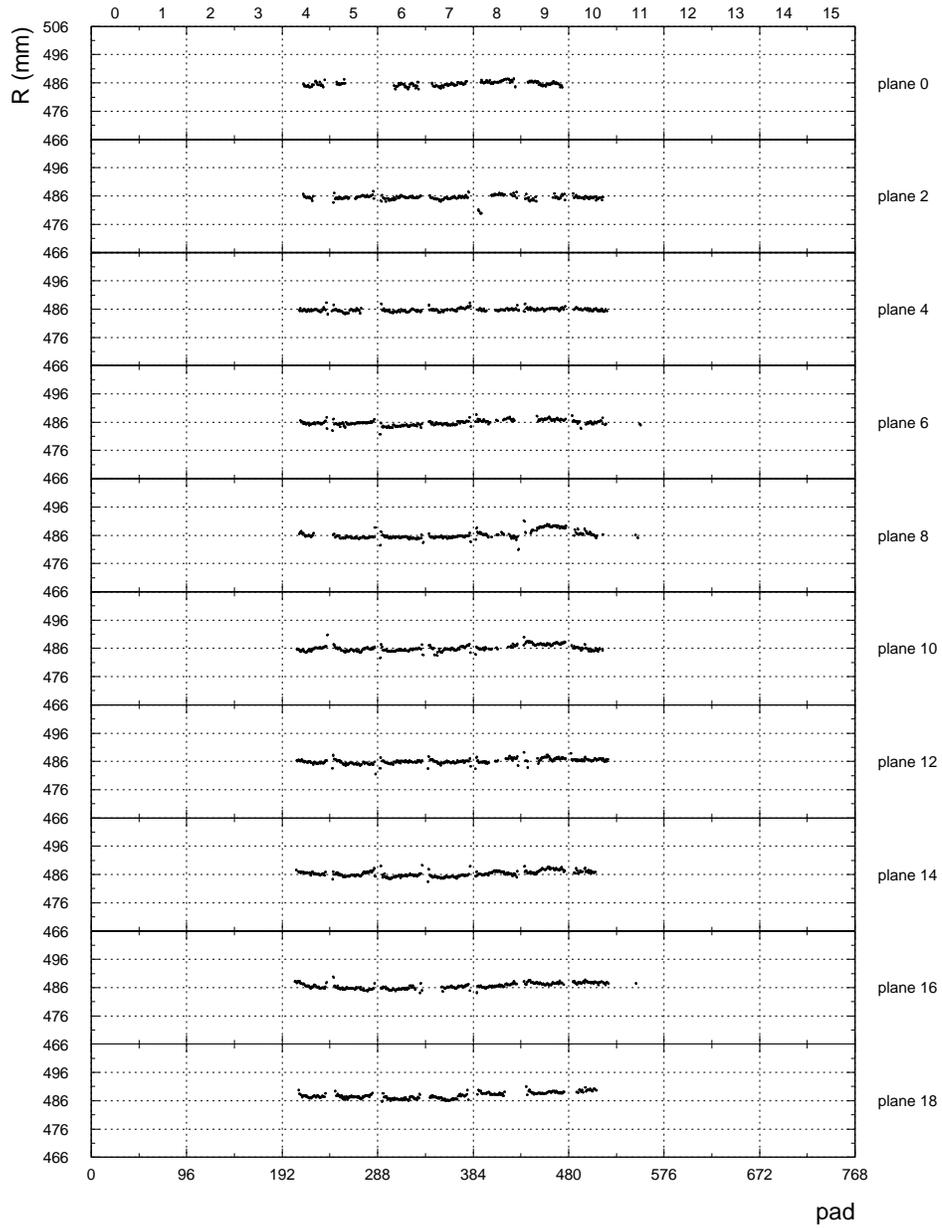}
\caption{
Reconstructed radius of the high voltage cylinder after calibration.}
\label{fig_ana-cylinder-e}
\vspace{1cm}
\end{figure}

The electrons knocked out of the high voltage cylinder by scattered laser 
light traverse the full drift volume before they hit a readout chamber. 
With proper calibration the reconstructed origin of the electrons 
should reproduce the radius of this cylinder $r$=486~mm. 
This constraint was used to calibrate the length of the drift path 
and the pad-to-pad time offsets. 
In a plot of the raw drift time measured for these electrons, 
shown in Fig.~\ref{fig_ana-cylinder-c}, 
the periodic structure in $\phi$, represented by the pad number, is the 
dominant feature. 
It is caused by the fact that the 16 readout chambers form a polygon 
around the otherwise cylindrical TPC and thus the drift path is longer 
for electrons arriving at the edge of a chamber. 
After accounting for this, the misalignment of the chambers, 
affecting the drift times via the path length and via the electric field, 
becomes visible as collective shifts or tilts of groups of 48 pads 
(Fig.~\ref{fig_ana-cylinder-e-raw}). 
Another effect manifest in the plot is a three-peak structure within 
each chamber caused by different capacities of the connections 
between the pads and the preamplifiers leading to differences in 
time offsets. The structure is related to three front-end-boards 
(group of 16 pads). 
Finally, individual time offsets were allowed for each front-end-board. 
The corrected data are shown in Fig.~\ref{fig_ana-cylinder-e}. 
It should be noted that a similar analysis can be performed using the 
edge of the time distribution collected for each pad in nuclear collision 
events. The calibration obtained this way matches more closely the 
experimental conditions during physics data taking. This way was actually 
used to calibrate the chamber positions and time offsets for the 
last CERES run in 2000. 
For a discussion of the resolutions achievable within these two methods 
see Section~\ref{sec_drift_monitoring}. 

\section{Determination of drift velocity and electric field corrections}
\label{sec_mobility}
The CERES TPC operated with an inhomogeneous electric field.
The exact knowledge of the electric field on one side and of the drift 
velocity for different field values on the other was necessary for track 
reconstruction. 
A calibration run consisting of 1000 laser events with seven laser rays 
in the special sector and with the photoelectron signal from the high 
voltage cylinder was used to determine the electron mobility function 
and the electric field corrections in absence of the magnetic field. 
This was done by minimizing radial distances of the 
reconstructed TPC hits from the expected track (and cylinder) positions. 
The expected track positions were calculated from the laser beam position 
before entering the TPC, measured with the position sensitive diodes, and 
the known mirror positions and angles, 
as described in Section~\ref{sec_entrance}. 
The 12 minimization parameters included three factors for electric field corrections 
at each end of the TPC, five parameters of the electron mobility curve, 
and the time offset.
(The electric field corrections were still needed to remove the remnants of 
the track curvature in $r$ vs. $z$, visible in Fig.~\ref{fig_laser329},  
after most of the effect was cured by taking into account the voltage 
dependence of the resistors in the divider chains and performing the 
three-dimensional potential calculation \cite{tpc-nim}). 

\begin{figure}[h]
\vspace{5mm} 
\hspace*{2cm}
\scalebox{0.65}{\includegraphics{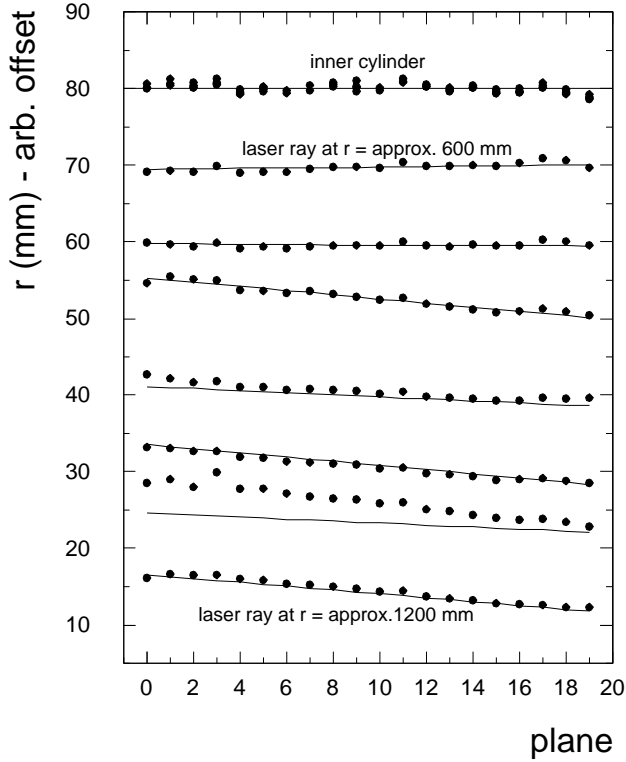}}\\
\vspace*{-5mm}
\caption{
Reconstructed laser tracks and the high voltage cylinder after the 
12-parameter field/mobility fit. 
The points and the lines represent calibrated measurements of the radial 
hit coordinate 
and the expected track positions based on the diode information, respectively, 
for the 20 $z$-planes of the TPC. }
\label{fig_mobility-fit87a}
\end{figure}

Fig.~\ref{fig_mobility-fit87a} shows the result of the minimization. 
The tracks were approximately parallel to the beam axis at $r$ = 60, 70, 
80, 90, 100, 110, and 120 cm; the radius of the high voltage cylinder was 
48.6 cm. 
For better visualisation in Fig.~\ref{fig_mobility-fit87a} the 
reconstructed tracks (points) were shifted closer to each other by arbitrary 
offsets; the same offsets were applied to the lines denoting the expected 
track positions.   
The difference between the points and the lines shows the residual 
distortions in the absolute scale. 
The track at $r$=110~cm, for which the mirror angle seems to deviate 
systematically from the assumed value, was excluded from the minimization. 

In a radial drift chamber, unlike in a typical TPC where the drift occurs 
at a fixed value of electric field, the whole mobility curve is needed for 
track reconstruction. 
The five-parameter mobility curve obtained from the fit differed 
significantly from results of a calculation performed with the drift 
simulation package Garfield/Magboltz~\cite{magboltz} 
(Fig.~15 in~\cite{tpc-nim}). 
The laser events with tracks at known positions were thus essential for 
understanding of the electron drift in the CERES TPC. 

\section{Drift velocity monitoring}
\label{sec_drift_monitoring}
\begin{figure}[h]
\hspace{5mm}\includegraphics[width=12cm]{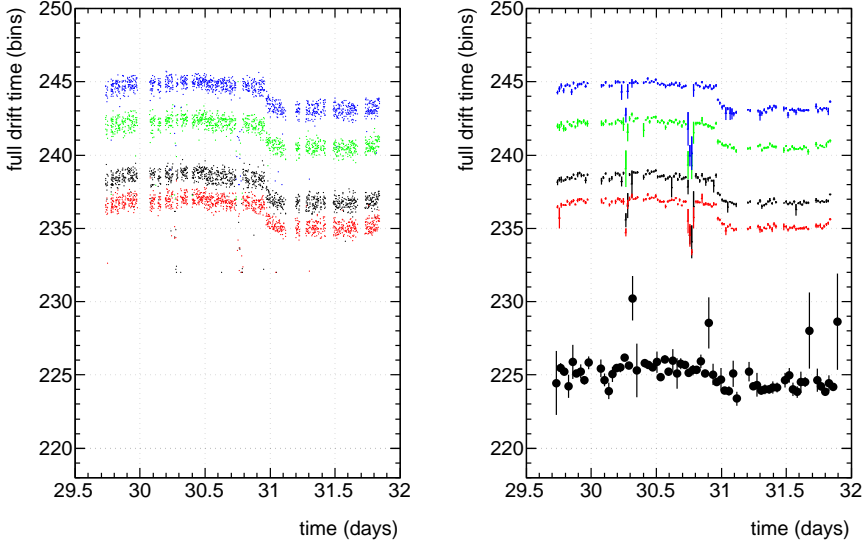}
\vspace{-5mm}
\caption{
Drift time of electrons from the high voltage cylinder to four groups of 
two pads, as a scatter plot (left panel) and a profile (lower four sets in 
the right panel), compared to the maximum drift time extracted from the 
edge of the time distribution of hits recorded in collision events 
(full dots in the right panel). 
The methods are consistently revealing the change in the drift velocity 
caused by a change in gas composition.
}
\label{fig_drif-velocity-edited}
\end{figure}

The drift time of the electrons knocked out from the high voltage cylinder 
in laser events can be used to monitor the changes in drift velocity 
with time. 
This mode of operation was tested during the Au+Pb run at 40~GeV per nucleon 
in 1999. 
Over several days of data taking in the pause between bursts the mirrors 
were flipped in and a laser event was generated. 
This event was becoming the first event on tape of the subsequent burst. 
The drift time of electrons registered in these events is shown in 
Fig.~\ref{fig_drif-velocity-edited}. 
In this figure, a period of two days is represented during which a distinct 
change in drift velocity occurred caused by the replacement of a gas filter. 
The drift times measured by four groups of two pads are represented 
as a scatter plot and as a profile in the left and right panel of 
Fig.~\ref{fig_drif-velocity-edited}, respectively. 
For comparison, filled dots in the right panel represents the full drift 
time extracted from the edge of the time distribution of hits recorded in 
nuclear collision events (for clarity shifted down by 20 timebins). 
The resolution of the laser method is 0.2 timebin (1 timebin = 294 ns) 
per pad per event. 
Properly combining one third of all pads (the laser illuminates only one 
side of the cylinder), i.e. 1/3 x 16 chambers x 20 planes x 
48 pads/chamber/plane = 5120 pads, should result in a resolution on 
the order of 0.003 timebins per event. 
This is to be compared to 0.9 timebin per event achievable with the 
edge method. 

While the laser method provides a much superior measurement of the total 
drift time in a single event, the low rate of laser events taken in the 
monitoring mode and the effort needed to maintain the system stable over 
many days make it less practicable than the other method. 
The final calibration of the drift velocity was performed with nuclear 
collision events using the edge of the radial hit distribution. 

\section{Summary}

The CERES TPC laser system was essential for verifying the electric 
field map and for understanding the electron drift in the inhomogeneous 
electric field, as well as for studying the detailed response of the readout 
chambers. 

It provided, furthermore, a convenient way to determine the gating grid 
transparency, to calibrate the chamber positions and the pad time offsets, 
and, by running the laser in parallel to production runs, 
to monitor the drift properties of the detector. 
Here, however, obtaining the analogue results from the recorded collision 
events was more practicable. 

The vigorous contribution of Wawrzyniec Prokopowicz to this project 
is gratefully acknowledged.
We thank Peter Liebold for building and improving the diode readout and 
the technicians Michael Marquardt and Joachim Weinert for their perfect 
work. 
Finally, we thank Hannes Wessels and Wilrid Dubitzky for carefully reading 
the manuscript and spotting many (hopefully all) errors and inconsistencies.

\end{document}